\documentclass[RNAAS]{aastex62}

\accepted{April 29, 2018}
\submitjournal{RNAAS}

\shorttitle{The quadruply lensed quasar ATLAS0259-1635}
\shortauthors{Schechter et al.}

\begin{document}

\title{Another quadruply lensed quasar from the VST-ATLAS survey}

\correspondingauthor{Paul L.\ Schechter}
\email{schech@mit.edu}

\author[0000-0002-5665-4172]{Paul L.\ Schechter}
\affil{MIT Kavli Institute 37-664G,\\
  77 Massachusetts Avenue, Cambridge, MA, 02138-4307, USA}\

\author[0000-0003-0930-5815]{T.\ Anguita}
\affil{Facultad de Ciencias Exactas, Universidad Andres Bello, Santiago, Chile}

\author[0000-0001-7779-9883]{Nicholas D.\ Morgan}
\affiliation{Staples High School, 70 North Avenue, Westport, CT 06680-2720, USA}

\author{M.\ Read}
\affil{Institute for Astronomy, University of Edinburgh, Royal Observatory, Edinburgh EH9 3HJ}

\author[0000-0003-2612-7926]{T.\ Shanks}
\affil{Durham University, Durham DH1 3LE, England}






\section{Search, Followup and Modelling}

Gravitational lenses producing four images of a quasar are prized for
four distinct purposes.  Time delay measurements yield direct
measurements of distances (Treu and Marshall 2016).  Flux ratio
anomalies are used both to measure surface stellar mass densities in
the lensing galaxies (Schechter et al 2014; Jimenez-Vicente et al
2015) and to measure the sizes of quasar continuum emitting regions
(Kochanek 2004; Pooley et al 2007).  They also provide multiple lines
of sight through intervening absorption line systems (Zahedy et al 2016).  But
quadruply lensed quasars are scarce, with only several dozen known.

\begin{figure}[b]
\plotone{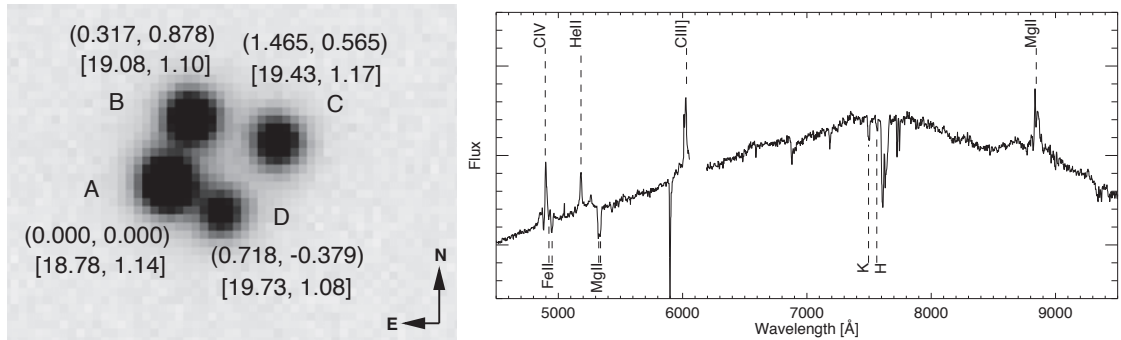}
\caption{Left: Sloan $i$ 120s exposure of ATLAS0259-1635  at $0\farcs111$ per pixel.  Coordinates
  are given in parentheses.  Sloan $i$ magnitudes and $g-i$ colors
  are given in brackets.  Right: Combined spectrum of images $A$ and $B$.}
\end{figure}

We report here the quadruple nature of the source WISE 025942.9-163543
as observed in the VST-ATLAS survey.  Followup spectra obtained with
IMACS (Dressler et al 2011) on the Baade 6.5 m telescope show two of the
four images to be components of a lensed quasar system with source
redshift z = 2.16, referred to heneforth at ATLAS 0259-1635

Our original search of the VST-ATLAS images for lensed quasars among
WISE sources with colors $W1-W2 > 0.7$ is described by Schechter et al
(2017).  Since then we have begun splitting sources into three
components, when possible, rather than just two.  Similar flux ratios
in the 5 filters are used to isolate lensed quasars.  In our original
search ATLAS 0259-1635 was split into two extended components, and
consequently, was passed over.  But when split into three components,
two were pointlike and one was extended, all with similar colors.
Figure 1 shows a direct image of ATLAS 0259-1635 taken in Sloan $i$ with
IMACS, along with positions for the four
images measured in arcseconds from the brightest component, image A,
at position $\alpha = 02^{\mathrm h}59^{\mathrm m}42\fs91,$ $\delta =
16\arcdeg35\arcmin43\farcs3$ on the ATLAS $r$ image.

PSF fitting photometry was obtained from 120 s $i$ and $g$ exposures
with IMACS using a star at $\alpha = 02^{\mathrm h}59^{\mathrm
  m}38\fs90s$, $\delta = -16\arcdeg36\arcmin22\farcs4$ as the template
for the four quasar images.  There was no convincing evidence of a lensing
galaxy in the residuals.  The $i$ magnitudes and $g-i$ colors (with
no adjustment for the difference in color between the template and the
quasar) are given in Figure 1.  Note that the quasar is quite red
for its redshift (Richards et al 2001).

Long slit spectroscopy was carried out with the 200l/mm grism on the f/2 camera
at PA $0^\circ$.  Spectra for images $A$ and $B$, shown in Figure 1,
have typical quasar emission lines at $z_s = 2.16$.  An absorption line
system is seen at $z_{abs} = 0.905$.  

Keeton's (2001) {\tt lensmodel} program was used to fit the positions
but not the fluxes of the four images to a singular isothermal sphere
with external shear. The best fitting model (with 7 free parameters)
gave a scatter of roughly $0\farcs01$ in the eight position
coordinates.  The galaxy is predicted to lie at (0.690, 0.319).  The
signed magnifications of images A, B, C and D are 10.5, -6.7, 6.4 and
-8.1, respectively.  The shear, $\gamma = 0.068$, points $19.7^\circ$
east of north.  The strength of the isothermal is $0\farcs733$.

\acknowledgments

We thank Dr. Magda Arnaboldi for the initial, excellent suggestion that
we might find it productive to collaborate.

%

\vspace{5mm}
\facilities{Magellan(IMACS)}

\end{document}